\documentclass[zpreprint,zbstpl]{./LaTeX/zeus/zeus_paper}
\usepackage[english]{babel}

\newcommand{\ZcoosysB}{%
The ZEUS coordinate system is a right-handed Cartesian system, with the $Z$
axis pointing in the proton beam direction, referred to as the ``forward
direction'', and the $X$ axis pointing left towards the centre of HERA.
The coordinate origin is at the nominal interaction point.\xspace}
\newcommand{\Zpsrap}{%
The pseudorapidity is defined as $\eta=-\ln\left(\tan\frac{\theta}{2}\right)$,
where the polar angle, $\theta$, is measured with respect to the proton beam
direction.\xspace}

\newcommand{\ZcoosysfnBeta}{\footnote{\ZcoosysB\Zpsrap}}
\newcommand{\Zdetdesc}{%
A detailed description of the ZEUS detector can be found 
elsewhere~\cite{zeus:1993:bluebook}. A brief outline of the 
components that are most relevant for this analysis is given
below.\xspace}
\newcommand{\Zctddesc}[1]{%
Charged particles are tracked in the central tracking detector (CTD)~\citeCTD,
which operates in a magnetic field of $1.43\Tesla$ provided by a thin 
superconducting coil. The CTD consists of 72~cylindrical drift chamber 
layers, organized in 9~superlayers covering the polar-angle#1 region 
\mbox{$15^\circ<\theta<164^\circ$}. The transverse-momentum resolution for
full-length tracks is
$\sigma(p_T)/p_T=0.0058p_T\oplus0.0065\oplus0.0014/p_T$ ($p_T$ in $\Gev$).}
\newcommand{\Zcaldesc}{%
The high-resolution uranium--scintillator calorimeter (CAL)~\citeCAL consists 
of three parts: the forward (FCAL), the barrel (BCAL) and the rear (RCAL)
calorimeters. Each part is subdivided transversely into towers and
longitudinally into one electromagnetic section (EMC) and either one (in RCAL)
or two (in BCAL and FCAL) hadronic sections (HAC). The smallest subdivision of
the calorimeter is called a cell.  The CAL energy resolutions, as measured under
test-beam conditions, are $\sigma(E)/E=0.18/\sqrt{E}$ for electrons and
$\sigma(E)/E=0.35/\sqrt{E}$ for hadrons ($E$ in $\Gev$).}
\chardef\usc=95
\chardef\til=126
\catcode`\@=11 % @ signs are now treated as letters
\DeclareRobustCommand\xdotspace{\futurelet\@let@token\@xdotspace}
\def\@xdotspace{%
  \ifx\@let@token.\else
  \ifx\@let@token\bgroup.\else
  \ifx\@let@token\egroup.\else
  \ifx\@let@token\/.\else
  \ifx\@let@token\ .\else
  \ifx\@let@token~.\else
  \ifx\@let@token!.\else
  \ifx\@let@token,.\else
  \ifx\@let@token:.\else
  \ifx\@let@token;.\else
  \ifx\@let@token?.\else
  \ifx\@let@token/.\else
  \ifx\@let@token'.\else
  \ifx\@let@token).\else
  \ifx\@let@token-.\else
  \ifx\@let@token\@xobeysp.\else
  \ifx\@let@token\space.\else
  \ifx\@let@token\@sptoken.\else
   .\space
   \fi\fi\fi\fi\fi\fi\fi\fi\fi\fi\fi\fi\fi\fi\fi\fi\fi\fi}
\catcode`\@=12 % @ signs are no longer letters

\newcommand{\stru}[2]{%
   \relax\ifmmode\hbox{\vrule height#1 depth#2 width0pt}%
   \else\vrule height#1 depth#2 width0pt\fi}
\newcommand{\Ronum}[1]{\uppercase\expandafter{\romannumeral#1}}
\newcommand{\ronum}[1]{\expandafter{\romannumeral#1}}
\DeclareRobustCommand{\LaTeXZ}{%
  \LaTeX\kern-.05em4\kern-.1em
  {\raisebox{-0.2ex}{$\scriptstyle\text{ZEUS}$}}\xspace}
\DeclareMathAlphabet{\mathbf}{OT1}{cmr}{bx}{sl}
\newcommand{\eVdist}{\kern-0.06667em}

\newcommand{\Gev}{{\text{Ge}\eVdist\text{V\/}}}

\newcommand{\gev}{{\,\text{Ge}\eVdist\text{V\/}}}

\newcommand{\Tesla}{\,\text{T}}

\newcommand{\slashfrac}[2]{%
  \raisebox{0.5ex}{\ensuremath #1}\kern-0.12em/\kern-0.08em
  \raisebox{-.8ex}{\ensuremath #2}}
\newcommand{\sqr}[3]{%
    {\vcenter{\hrule height.#3ex\hbox{\vrule width.#2ex height#1ex
     \kern#1ex\vrule width.#3ex}\hrule height.#2ex}}}

\catcode`\@=11 % @ signs are now treated as letters
\newcommand{\parenbar}{\mathpalette\p@renb@r}
\def\p@renb@r#1#2{\vbox{%
  \ifx#1\scriptscriptstyle \dimen@.7em\dimen@ii.2em\else
  \ifx#1\scriptstyle \dimen@.8em\dimen@ii.25em\else
  \dimen@1em\dimen@ii.4em\fi\fi \offinterlineskip
  \ialign{\hfill##\hfill\cr
    \vbox{\hrule width\dimen@ii}\cr
    \noalign{\vskip-.3ex}%
    \hbox to\dimen@{$\mathchar300\hfil\mathchar301$}\cr
    \noalign{\vskip-.3ex}%
    $#1#2$\cr}}}
\catcode`\@=12 % @ signs are no longer letters

\newcommand{\IP}{{\rm I$\kern-0.01667em$P}\xspace}

\mathchardef\qsm=63
\mathchardef\pls=43
\mathchardef\mns=512
\mathchardef\plm=518
\mathchardef\eql=61
\mathchardef\smallleft=300
\mathchardef\smallright=301
\mathchardef\les=316
\mathchardef\gre=318
\mathchardef\leq=532
\mathchardef\grq=533
\catcode`\@=11 % @ signs are now treated as letters
\newcounter{pict@width}
\newcounter{pict@height}
\newlength{\pict@scale}
\newcommand{\psfigadd}[4]{%
\setcounter{pict@width}{1*\ratio{#2+\pict@scale/2}{\pict@scale}}
\setcounter{pict@height}{1*\ratio{#3+\pict@scale/2}{\pict@scale}}
\setlength{\unitlength}{\pict@scale}
\hbox to #2{\hspace{-\fill}\begin{picture}(\thepict@width,\thepict@height)
\put(0,0){\psfig{figure=#1,width=#2,height=#3,clip=}}
\SetScale{0.283466457}
\SetWidth{1.763889}
{#4}
\end{picture}}
}
\newcounter{pict@widthfst}
\newcounter{pict@widthscd}
\newcounter{pict@widthtot}
\newcommand{\psfigaddtwo}[7]{%
\setcounter{pict@widthfst}{1*\ratio{#2+\pict@scale/2}{\pict@scale}}
\setcounter{pict@widthscd}{1*\ratio{#2+#4+\pict@scale/2}{\pict@scale}}
\setcounter{pict@widthtot}{1*\ratio{#2+#4+#6+\pict@scale/2}{\pict@scale}}
\setcounter{pict@height}{1*\ratio{#3+\pict@scale/2}{\pict@scale}}
\setlength{\unitlength}{\pict@scale}
\hbox{\hspace{-\fill}\begin{picture}(\thepict@widthtot,\thepict@height)
\put(0,0){\psfig{figure=#1,width=#2,height=#3,clip=}}
\put(\thepict@widthscd,0){\psfig{figure=#5,width=#6,height=#3,clip=}}
\SetScale{0.283466457}
\SetWidth{1.763889}
{#7}
\end{picture}}
}
\newcommand{\psfigror}[4]{%
\setcounter{pict@width}{1*\ratio{#2+\pict@scale/2}{\pict@scale}}
\setcounter{pict@height}{1*\ratio{#3+\pict@scale/2}{\pict@scale}}
\setlength{\unitlength}{\pict@scale}
\hbox{\begin{picture}(\thepict@width,\thepict@height)
\put(0,\thepict@height){\psfig{figure=#1,width=#3,height=#2,clip=,angle=270}}
\SetScale{0.283466457}
\SetWidth{1.763889}
{#4}
\end{picture}}
}
\newcommand{\psfigrol}[4]{%
\setcounter{pict@width}{1*\ratio{#2+\pict@scale/2}{\pict@scale}}
\setcounter{pict@height}{1*\ratio{#3+\pict@scale/2}{\pict@scale}}
\setlength{\unitlength}{\pict@scale}
\hbox{\begin{picture}(\thepict@width,\thepict@height)
\put(0,0){\psfig{figure=#1,width=#3,height=#2,clip=,angle=90}}
\SetScale{0.283466457}
\SetWidth{1.763889}
{#4}
\end{picture}}
}
\catcode`\@=12 % @ signs are no longer letters
\newlength\listtextwidth

\catcode`\@=11 % @ signs are now treated as letters
\newlength{\@tabfninsert}
\newlength{\@tabfnwidth}
\newcommand{\tabfootnote}[2]{%
  \setlength{\@tabfninsert}{0.8em}
  \setlength{\@tabfnwidth}{\textwidth}
  \addtolength{\@tabfnwidth}{-\@tabfninsert}
  \addtolength{\@tabfnwidth}{-0.4em}
  \noindent\makebox[\@tabfninsert][r]{\footnotesize$^{#1}$\hfil}\hfill%
  \parbox[t]{\@tabfnwidth}{\footnotesize #2\hfill}}
\catcode`\@=12 % @ signs are no longer letters

\def\citeCTD{{\cite{%
nim:a279:290,*npps:b32:181,*nim:a338:254%
}}\xspace}
\def\citeCAL{{\cite{%
nim:a309:77,*nim:a309:101,*nim:a321:356,*nim:a336:23%
}}\xspace}

\begin{document}
%------------------------------------------------------------------------------
%       Title sheet
%------------------------------------------------------------------------------
\title{
\vspace{-5cm}
\begin{flushleft} {\normalsize \tt DESY 03-015}\\ \vspace{-.25cm}{\normalsize \tt February 2003} \end{flushleft}
\vspace{2cm}
Dijet angular distributions \\
in photoproduction of charm at HERA
}                                                       
                    
\author{ZEUS Collaboration}
\draftversion{Post-reading}
\date{ 23 \ January 2003}
 
\abstract{
Dijet angular distributions of photoproduction events in which a $D^{*\pm}$ meson 
is produced in association with one of two energetic jets have been measured 
with the ZEUS detector at HERA, using an integrated luminosity of 120 pb$^{-1}$. 
Differential cross sections as a function of the angle between the charm-jet 
and the proton-beam direction in the dijet rest frame  have been 
measured for samples enriched in direct or resolved photon events. The results 
are compared with predictions from leading-order parton-shower Monte Carlo 
models and with next-to-leading-order QCD calculations. The angular 
distributions show clear evidence for the existence of charm originating from the 
photon.
}
 
\makezeustitle

\def\3{\ss}                                                                                        
\pagenumbering{Roman}                                                                              
                                    % this "%"s are for cosmetics only                             
%\begin{document}                                                                                   
                                                   %                                               
\begin{center}                                                                                     
{                      \Large  The ZEUS Collaboration              }                               
\end{center}                                                                                       
  S.~Chekanov,                                                                                     
  M.~Derrick,                                                                                      
  D.~Krakauer,                                                                                     
  J.H.~Loizides$^{   1}$,                                                                          
  S.~Magill,                                                                                       
  B.~Musgrave,                                                                                     
  J.~Repond,                                                                                       
  R.~Yoshida\\                                                                                     
 {\it Argonne National Laboratory, Argonne, Illinois 60439-4815}~$^{n}$                            
\par \filbreak                                                                                     
  M.C.K.~Mattingly \\                                                                              
 {\it Andrews University, Berrien Springs, Michigan 49104-0380}                                    
\par \filbreak                                                                                     
  P.~Antonioli,                                                                                    
  G.~Bari,                                                                                         
  M.~Basile,                                                                                       
  L.~Bellagamba,                                                                                   
  D.~Boscherini,                                                                                   
  A.~Bruni,                                                                                        
  G.~Bruni,                                                                                        
  G.~Cara~Romeo,                                                                                   
  L.~Cifarelli,                                                                                    
  F.~Cindolo,                                                                                      
  A.~Contin,                                                                                       
  M.~Corradi,                                                                                      
  S.~De~Pasquale,                                                                                  
  P.~Giusti,                                                                                       
  G.~Iacobucci,                                                                                    
  A.~Margotti,                                                                                     
  R.~Nania,                                                                                        
  F.~Palmonari,                                                                                    
  A.~Pesci,                                                                                        
  G.~Sartorelli,                                                                                   
  A.~Zichichi  \\                                                                                  
  {\it University and INFN Bologna, Bologna, Italy}~$^{e}$
% ...
\newpage                                                                                           
$^{\    1}$ also affiliated with University College London \\                                      
% ...
                                                           %                                       
                                                           %                                       
% \par         % if index listing & table fit to 1 page, put gap here                              
\newpage   % alternatively: go to newpage, if page is too small                                    
                                                           %                                       
% \institute_references_start    % do not touch or move this line !                                
                                                           %                                       
\begin{tabular}[h]{rp{14cm}}                                                                       
$^{a}$ &  supported by the Natural Sciences and Engineering Research                               
          Council of Canada (NSERC) \\                                                             
$^{b}$ &  supported by the German Federal Ministry for Education and                               
          Research (BMBF), under contract numbers HZ1GUA 2, HZ1GUB 0, HZ1PDA 5, HZ1VFA 5\\         
$^{c}$ &  supported by the MINERVA Gesellschaft f\"ur Forschung GmbH, the                          
          Israel Science Foundation, the U.S.-Israel Binational Science                            
          Foundation and the Benozyio Center                                                       
          for High Energy Physics\\                                                                
$^{d}$ &  supported by the German-Israeli Foundation and the Israel Science                        
          Foundation\\                                                                             
$^{e}$ &  supported by the Italian National Institute for Nuclear Physics (INFN) \\                
$^{f}$ &  supported by the Japanese Ministry of Education, Science and                             
          Culture (the Monbusho) and its grants for Scientific Research\\                          
$^{g}$ &  supported by the Korean Ministry of Education and Korea Science                          
          and Engineering Foundation\\                                                             
$^{h}$ &  supported by the Netherlands Foundation for Research on Matter (FOM)\\                   
$^{i}$ &  supported by the Polish State Committee for Scientific Research,                         
          grant no. 620/E-77/SPUB-M/DESY/P-03/DZ 247/2000-2002\\                                   
$^{j}$ &  partially supported by the German Federal Ministry for Education                         
          and Research (BMBF)\\                                                                    
$^{k}$ &  supported by the Fund for Fundamental Research of Russian Ministry                       
          for Science and Edu\-cation and by the German Federal Ministry for                       
          Education and Research (BMBF)\\                                                          
$^{l}$ &  supported by the Spanish Ministry of Education and Science                               
          through funds provided by CICYT\\                                                        
$^{m}$ &  supported by the Particle Physics and Astronomy Research Council, UK\\                   
$^{n}$ &  supported by the US Department of Energy\\                                               
$^{o}$ &  supported by the US National Science Foundation\\                                        
$^{p}$ &  supported by the Polish State Committee for Scientific Research,                         
          grant no. 112/E-356/SPUB-M/DESY/P-03/DZ 301/2000-2002, 2 P03B 13922\\                    
$^{q}$ &  supported by the Polish State Committee for Scientific Research,                         
          grant no. 115/E-343/SPUB-M/DESY/P-03/DZ 121/2001-2002, 2 P03B 07022\\                    
\end{tabular}                                                                                      
                                                           %                                       
% \institute_references_end     % do not touch or move this line !                                 
                                                           %                                       
%\end{document}                                                                                     

%------------------------------------------------------------------------------
%       Text
%------------------------------------------------------------------------------
\pagenumbering{arabic} 
\pagestyle{plain}
\def\ds{D^{\ast}}
\def\dspm{{\ds}^{\pm}}
\def\xgo{x_\gamma^{\rm obs}}
\def\xpo{x_p^{\rm obs}}
\def\ETJT{E_T^{\rm jet}}
\def\WJB{W_{\textrm{JB}}}
% ----------------------------------------------------------------------------
%       Introduction
% ----------------------------------------------------------------------------
\section{Introduction}
\label{sec-int}
 
High-energy collisions at the HERA $ep$ collider between a quasi-real photon 
and a proton provide an effective source of photoproduction processes. Jets with high 
transverse energy and/or charm ($c$) quarks produced in such processes can be 
described within quantum chromodynamics (QCD) in two sub-classes: direct 
processes are those in which the photon couples as a point-like object in the 
hard scattering; resolved processes are those in which the photon acts as a 
source of incoming partons, one of which participates in the hard interaction. 
Both processes can lead to two jets in the final state. Samples enriched in 
direct and resolved photon events can be identified using the variable 
$\xgo$~\cite{pl:b348:665}, which is the fraction of the photon's momentum 
contributing to the production of the two jets.
 
Inclusive cross sections for photoproduction of $D^{*\pm}(2010)$
mesons as well as cross sections for ``charm dijet" events, in which
the $D^*$ is observed in events with two energetic jets, have been
previously reported~\cite{epj:c6:67}. Differential cross sections of
the $D^*$ and associated dijet system are larger than
next-to-leading-order (NLO) QCD
predictions~\cite{np:b412:225,*pl:b348:633,*np:b454:3} at low $\xgo$, but are in
agreement at high $\xgo$. The data were also compared to predictions
of leading-logarithmic parton-shower Monte Carlo (MC) models.
According to these comparisons, about 60\% of the events can be
attributed to the direct photon-gluon-fusion (PGF) process $\gamma
g\to c\bar c$, illustrated in Fig.~1a). The MC models predict that
most of the resolved photon events come from charm excitation of the
photon (Figs.~1c) and 1d)) rather than from the $g g \to c \bar c$
process (Fig.~1b)). 
The aim of this analysis is to determine the
dominant mechanisms for charm dijet photoproduction in both direct and
resolved photon processes.

Measuring the angular distribution of the
outgoing jets allows the dominant subprocesses to be determined and
the MC predictions to be tested, as was done previously~\cite{pl:b384:401,*epj:c23:615}
for inclusive dijet events. This study showed that the differential
cross-section $d\sigma /d|\cos\theta^*|$, where $\theta^*$ is the
angle between the jet-jet axis and the proton beam direction in the
dijet rest frame, is sensitive to the spin of the propagator in the
hard subprocess. In direct photon processes, in which the propagator
in the leading-order (LO) QCD diagrams is a quark, the differential
cross section rises slowly towards high $|\cos \theta^*|$ values
($d\sigma /d|\cos\theta^*|\approx (1-|\cos\theta^*|)^{-1}$). In
resolved photon processes, the gluon propagator is allowed at LO and 
 dominates over the quark propagator due to the stronger gluon-gluon
coupling compared to the quark-gluon coupling. In this case 
the cross section rises steeply when $|\cos \theta^*|$
increases ($d\sigma /d|\cos\theta^*|\approx (1-|\cos\theta^*|)^{-2}$).
Similar results have been reported in photon-photon
collisions~\cite{epj:c10:547,*hep-ex-0301013}.
 
If most of the resolved-photon charm dijet events are produced as a
result of charm from the photon, a gluon-exchange contribution,
as seen in Fig.~1d), should dominate. This results in a steep rise of
the cross section towards high $|\cos\theta^*|$ values. The other
diagrams of Fig.~1 involve quark exchange and thus should not show
such a sharp rise.  If one of the jets is explicitly tagged as a charm
jet, the sign of $\cos \theta^* $ can be defined. If the charm
originates from the photon, the charm jet generally lies in the
photon hemisphere.
 
% ----------------------------------------------------------------------------
%       Experimental set-up
% ----------------------------------------------------------------------------
\section{Experimental conditions}
\label{sec-exp}
 
The analysis was performed using data collected with the ZEUS detector at HERA 
during 1996 -- 2000. In this period, HERA collided electrons or positrons with 
energy $E_e~=~27.5\gev$ and protons with energy $E_p =820\gev$ (1996 -- 1997) or 
$E_p =920\gev$ (1998 -- 2000), corresponding                to integrated luminosities of 
$38.6\pm 0.6$ and $81.9\pm 1.8~{\rm pb^{-1}}$ and to                            
centre-of-mass energies $\sqrt{s} = 300\gev$ and $\sqrt{s} = 318\gev$, respectively.
                                                             This data sample 
is about a factor of three larger than that used for the previous charm dijet 
analysis~\cite{epj:c6:67}.
 
\Zdetdesc
 
\Zctddesc\ZcoosysfnBeta
 
\Zcaldesc
 
%see draft for other comments on changes to predefined commands
 
%LUMI
 
The luminosity was measured from the rate of the bremsstrahlung process 
$e^+p~\rightarrow~e^+\gamma p$, where the photon was measured in a lead-scintillator 
%calorimeter~\cite{desy-92-066,*zfp:c63:391,*desy-01-141} placed in the HERA tunnel at $Z=-107~{\rm m}$.
calorimeter~\cite{desy-92-066,*zfp:c63:391,*acpp:b32:2025} placed in the HERA tunnel at $Z=-107~{\rm m}$.
%acpp:b32:2025
 
% ----------------------------------------------------------------------------
\section{Event selection}
\label{sec-evt}
                  
Photoproduction events were selected with a three-level 
%trigger~\cite{zeus:1993:bluebook,cpaper:ichep92:DAQ}. The inclusive 
 trigger~\cite{zeus:1993:bluebook,proc:chep:1992:222}. The inclusive 
photoproduction sample was defined by requiring a reconstructed vertex and no 
scattered electron or positron found in the CAL, thus restricting the photon 
virtuality, $Q^2$, to be below 1\gev$^2$, with median 
$Q^2\approx~3\cdot~10^{-4}$\gev$^2$. The photon-proton centre-of-mass energy, $W$, 
was restricted to the range $130~<~W~<~280\gev$. The latter was measured 
using the Jacquet-Blondel~\cite{proc:epfacility:1979:391} estimator 
$\WJB~=~\sqrt{4 y_{\rm JB}E_e E_p}$, where 
$y_{\rm JB}~=~{\Sigma_{i}(E_i-p_{Z,i})}/2E_e$,       the sum runs over all CAL 
cells and $p_{Z,i}$ is the $Z$ component of the momentum vector assigned to each 
cell of energy $E_i$. Jets were reconstructed with the $k_T$ cluster 
algorithm~\cite{np:b406:187} in its longitudinally invariant inclusive 
mode~\cite{pr:d48:3160}. The events were required to have at least two 
jets\footnote{The fraction of events with more than two jets is $11\%$.} with 
pseudorapidity $|\eta^{\rm jet}| < 2.4$ and transverse energy 
$E_T^{\rm jet} > 5\gev$. The measured jet energies as well as $\WJB$ were 
corrected for energy losses in inactive material in front of the CAL, using the 
MC simulation.
 
The $D^{*}$ mesons were reconstructed using the mass-difference technique 
applied to the decay chain\footnote{Throughout this article,
 $D^0$ refers to both $D^0$ and $\bar D^0$.} $D^{*\pm}\to
D^0\pi^{\pm}_S\to K^{\mp}\pi^{\pm}\pi^{\pm}_S$. Tracks in the CTD  
with opposite charges and transverse momenta $p_T > 0.5\gev$ were combined in 
pairs to form $D^0$ candidates. Kaon and pion masses were assumed in turn for 
each track to calculate the pair invariant mass, $M(K\pi)$. A third track, 
$\pi_S$, assumed to be the ``soft pion" from the $D^*$ decay, with 
$p_T > 0.15\gev$ and a charge opposite to the kaon, was added to form a $D^{*}$ 
candidate. Events with a mass difference $\Delta M = M(K\pi\pi_s) - M(K\pi)$ in the 
range $0.1435 < \Delta M < 0.1475\gev$ around the nominal value~\cite{epj:c15:1} 
and the range $1.81 < M(K\pi) < 1.92\gev$ around the $D^0$ mass were
called $D^{*}$ candidates. To suppress 
combinatorial background, a cut $p_T^{\ds} / E_T^{\theta > 10^{\circ}} > 0.15$ 
was applied~\cite{epj:c6:67}, where $ E_T^{\theta > 10^{\circ}}$ is the transverse 
energy measured in the CAL outside a cone of $\theta = 10^\circ $ in the forward 
direction. The reconstructed $D^*$ mesons were required to have 
$p_T^{D^*} > 3\gev$ and pseudorapidity in the range $|\eta^{D^*}|< 1.5$. 
 
These cuts ensure that the events lie in a well understood acceptance region of 
the detector.
 
%%%%% Mjj, etabar, cos(theta*) cuts
 
\section{Jet kinematic variables}
 
Samples enriched in direct and resolved photon events were separated
by a selection on the variable
 
\begin{equation}
\xgo = \frac{\Sigma_{\rm jets}\left(\ETJT\ e^{-\eta^{\rm jet}}\right)}{2 y E_e}, \nonumber
\end{equation}
 
where $y E_e$ is the initial photon energy and                                 
      the sum is over the two jets with the highest $\ETJT$.                 
%                                                            The fraction of 
%the electron or positron beam energy carried by the photon, $y$, was measured 
%using the Jacquet-Blondel estimator $y_{\rm JB}$.                     
                                                   The selection of 
$\xgo > 0.75$ and $\xgo < 0.75$ yields samples enriched in direct and 
resolved photon processes, respectively.
 
A complementary variable is 
 
\begin{equation}
\xpo = \frac{\Sigma_{\rm jets}\left(\ETJT\ e^{ \eta^{\rm jet}}\right)}{2 E_p}, \nonumber
\end{equation}
 
which is the fraction of the proton's momentum contributing to the production of 
the two jets.
 
The dijet scattering angle, $\theta^*$, is reconstructed using 
 
\begin{equation}
\cos \theta^* = \tanh\left(\frac{\eta^{\rm jet1}-\eta^{\rm jet2}} {2}\right).                   
\end{equation}
 
In the simple case in which two jets are back-to-back in the transverse plane and 
have equal transverse energies, the dijet invariant mass is given by 
$M_{\rm jj} = 2 E_T^{\rm jet}/\sqrt { 1 - |\cos \theta^*|^2 }$. Therefore, for a 
given $M_{\rm jj}$, events with high values of $|\cos \theta^*|$ have lower 
$E_T^{\rm jet}$. In order to study the $|\cos\theta^*|$ distribution up to 
$|\cos \theta^*| = 0.83$ without bias from the $E_T^{\rm jet}$ cut, $M_{\rm jj}$ 
was required to be above $18\gev$.
 
A cut on the average longitudinal boost, 
$\bar\eta = (\eta^{\rm jet1} + \eta^{\rm jet2})/2$, of $|\bar\eta| < 0.7$ was  
applied. This selection limits $\eta^{\rm jet}$ to $|\eta^{\rm jet}| < 1.9$ and 
removes the bias caused by the explicit cuts on $\eta^{\rm jet}$\cite{pl:b384:401}. 
It also reduces the bias caused by the cut on $|\eta^{D^*}| < 1.5$ while retaining 
a sufficiently large number of events. Monte Carlo studies show that the residual distortion due 
to the $|\eta^{D^*}|$ cut is small and    confined to the extreme bins of the 
$\cos \theta^* $ distribution. 
 
These cuts ensure that any features seen in the measured distributions can be 
attributed to the dynamics of the hard scattering processes.

% ----------------------------------------------------------------------------
\section{Models and QCD calculations}
\label{sec-mc}
 
The MC simulation programs PYTHIA 6.156~\cite{cpc:135:238} and
HERWIG 6.301~\cite{cpc:67:465,*jhep:0101:010,*hep-ph-0107071} were used to model the final states. The PYTHIA and 
HERWIG simulations use on-shell LO matrix elements for charm photoproduction 
processes. Higher-order QCD effects are simulated in the leading-logarithmic 
approximation with initial- and final-state radiation obeying DGLAP 
evolution~\cite{sovjnp:15:438,*sovjnp:20:94,*np:b126:298,*jetp:46:641}. Coherence 
effects from soft-gluon interference are included. The parton density functions 
(PDF) CTEQ5L~\cite{epj:c12:375} for the proton and GRV-G LO~\cite{pr:d46:1973} for 
the photon were used. The LO direct and resolved photon processes were generated 
proportionally to their predicted MC cross sections, using charm- and beauty-quark 
masses of $m_c = 1.5\gev$ and $m_b = 4.75\gev$, respectively. Fragmentation into 
hadrons is simulated in HERWIG with a cluster algorithm~\cite{np:b238:492} and in 
PYTHIA with the Lund string model~\cite{prep:97:31}.
    
Samples of MC events larger than the dataset were produced. 
To calculate the acceptances and to 
estimate     hadronisation effects, the events were passed through the 
GEANT 3.13-based~\cite{manual:cern-DD/EE/84-1} simulation of the ZEUS detector and trigger. 
They were reconstructed and analysed by the same program chain as the data.
Samples corresponding to different data taking conditions were
generated in proportion to 
their luminosities. For PYTHIA, in addition to the $D^*$ decay chain used for this analysis,
background events  that arise from other $D^{*\pm}$ decay modes or similar decay modes
of other charm mesons  were also simulated.
    
%The MC event generator CASCADE 1.00/09~\cite{hep-ph-0109146}  simulates heavy-quark 
The MC event generator CASCADE
1.00/09~\cite{epj:c19:351,*cpc:143:100,*jpg:28:971} simulates heavy-quark 
photoproduction in the framework of the
   semi-hard                                                                 
   or $k_t$-factorisation 
  approach~\cite{prep:100:1,*prep:189:267,*sovjnp:53:657,*sovjnp:54:867,
                               *pl:b242:97,*np:b366:135,*np:b360:3,*np:b386:215}.
          The      matrix element used in CASCADE is the off-shell LO 
PGF process. 
%Important partial contributions, which are             
%of        NLO and even next-to-next-to-leading-order nature        in the collinear
%(on-shell) approach, are consistently included in     $k_t$-factorisation due to the
%off-shellness of the gluons entering the PGF process~\cite{pan:64:120}.
             The CASCADE initial-state radiation is based on CCFM 
evolution~\cite{np:b296:49,*pl:b234:339,*np:b336:18,*np:b445:49}, which includes in the 
perturbative expansion the $\ln (1/x)$ terms in addition to the $\ln Q^2$ terms used in 
DGLAP evolution. To simulate final-state radiation, CASCADE uses PYTHIA 6.1 and the 
fragmentation into hadrons is simulated with the Lund string model. The cross section 
is calculated by convoluting the off-shell PGF matrix element with the unintegrated 
gluon density of the proton obtained from the CCFM fit to the HERA $F_2$ 
%data~\cite{np:b470:3}.                                                            
%data~\cite{epj:c19:351}.                                                            
data, by fixing most of the                                                         
 free parameters~\cite{epj:c19:351,*cpc:143:100,*jpg:28:971}.
%A charm quark mass of $m_c = 1.5\gev$ was used in this paper.                
                                      Although the CASCADE matrix element 
corresponds to the off-shell                 
             PGF direct photon process only (Fig.~1a)), resolved photon processes 
%are  reproduced by the CCFM initial-state radiation~\cite{pl:b491:111,*hep-ph-0203025}. 
are reproduced by the CCFM initial-state radiation~\cite{pl:b491:111,*epj:c24:425}. 
    
The NLO QCD calculations of differential cross sections for photoproduction of charm 
dijet events in the HERA kinematic region are available~\cite{np:b412:225,*pl:b348:633,*np:b454:3} 
in the fixed-order (FO)  scheme. The PDF parameterisations used were 
CTEQ5M1~\cite{epj:c12:375} for the proton and AFGHO~\cite{ZfP:c64:621} for the photon. 
The factorisation scales of the photon and proton PDFs, $\mu_F$, and the 
renormalisation scale, $\mu_R$, used for the calculation were set to 
$\mu_F =\mu_R = m_T\equiv\sqrt{ m_{c}^{2} + \langle  p_T^{2} \rangle } $, where           
                          $\langle p_T^{2} \rangle$ was set to the average $p_T^2$ of 
the charm quark and antiquark. The charm fragmentation into $D^*$ was performed using 
the Peterson fragmentation function~\cite{pr:d27:105} with an $\epsilon$ parameter of 
0.035~\cite{np:b565:245}.
 
In all cases, the fraction of $c$ quarks fragmenting into a $D^*$ was assumed to be 
0.235~\cite{hep-ex-9912064} and
 a charm quark mass of $m_c = 1.5\gev$ was used.                              
 
% ----------------------------------------------------------------------------
\section{Results}
\label{sec-res}

The $\Delta M$ distribution for dijet events in the $D^0$ signal
region shows a clear $D^*$ signal. The analysis is based on $1092\pm
43$ $D^{*\pm}$ mesons found in the $0.1435 < \Delta M < 0.1475\gev$
region over a background of
328  events.  The signal has similar characteristics as that in the
previous ZEUS publication~\cite{epj:c6:67} except that the signal to background
ratio has improved by a factor of three due to the tighter cuts (see
Sections 3 and 4) used here.
The background was determined from the $\Delta M$ distribution for wrong-charge combinations, 
where the tracks forming     $D^0$ candidates had  the same charge and the $\pi_S$ had 
the opposite charge.                                       
 
The number of events in each bin of the measured variables was extracted by performing a 
bin-by-bin wrong-charge background subtraction. To obtain differential cross sections, 
each value was then multiplied by a correction factor proportional to the 
ratio of generated to reconstructed    events  from the PYTHIA MC simulation. The 
measured cross sections are the luminosity-weighted average of the cross sections at the 
centre-of-mass energies $\sqrt{s} = 300$~GeV and $\sqrt{s} = 318$~GeV.
 
%A detailed study of possible sources of systematic uncertainties was carried out.
%The largest contributions were due to the variation of the cuts on $W$ and from 
%correcting the data with HERWIG rather than with PYTHIA. Shifts in the CAL energy scale 
%of $\pm 3\%$ are correlated between bins and are therefore not included in the overall 
%uncertainty, which is obtained by adding the remaining uncertainties in quadrature. The 
%statistical uncertainties dominate.

The systematic uncertainties were determined by adding the
contributions from several sources in quadrature. The 
largest contributions were associated with the cuts on $W$ and with
the difference between the correction factors evaluated using HERWIG
rather than PYTHIA. The uncertainties due to the knowledge of the CAL
energy scale ($\pm 3\%$) are highly correlated between bins and are
therefore shown separately. Statistical uncertainties dominate over
systematic ones in most bins. The measured cross sections and their
uncertainties are given in Tables 1-4.

The differential cross section  as a function of $\xgo$ is shown in Fig.~2. The peak 
at high values of $\xgo$ indicates a large contribution 
from direct photon processes, but there is also a sizeable contribution from resolved 
photon processes at lower $\xgo$ values. Figure~3 shows                               
the differential cross section  as a function of $\xpo$.                             
                      The $\xpo$ range of the 
data is concentrated in the region $0.0055 < \xpo < 0.044$, where the proton PDFs are well 
determined.
 
Figure 4 shows the differential cross sections as a function of $|\cos\theta^*|$ 
separately for the resolved-enriched ($\xgo < 0.75$) and direct-enriched ($\xgo > 0.75$) 
samples. The cross section for the sample enriched in resolved photons exhibits a more 
rapid rise towards high values of $|\cos\theta^*|$ than does the cross section for 
the sample enriched in direct photons.          
Consequently, the LO subprocess $gg\to c\bar c$ (Fig.~1b)), with
$q$-exchange in the $t$ channel, cannot 
be the dominant resolved photon process for charm dijet events. 
                              This observation suggests a large gluon-exchange 
contribution originating from a charm-excitation process.                        
 
The $|\cos\theta^*|$ distributions of Fig.~4 are similar in shape to the previously 
reported dijet angular distributions~\cite{pl:b384:401}, which did not require the 
presence of charm. In those analyses, only the absolute value of $\cos\theta^*$ was determined. 
In the present study, the two jets were distinguished by associating the $D^*$ meson to 
the closest jet 
in $\eta - \phi$ space. The associated jet is defined to be the jet with the smallest
$R_i=\sqrt{(\eta^{{\rm jet},i}-\eta^{D^*})^2 + (\phi^{{\rm
jet},i}-\phi^{D^*})^2}$ ;  
$(i=1,2)$ and with $R < 1$, where $\phi^{\rm jet}$ ($\phi^{D^*}$) is the azimuthal 
angle of the jet ($D^*$) in the laboratory frame. Calling this ``$D^*$ jet" 
jet~1 in Eq.~(1), the rise of $d\sigma /d\cos\theta^*$ can be studied separately 
for the photon and proton directions. Figure~5 shows the differential cross sections 
as a function of $\cos\theta^*$ for the resolved- and direct-enriched samples. Events 
that did not satisfy the requirement $R < 1$ for at least one of the two jets 
($8.7\%$ for $\xgo < 0.75$ and $1.1\%$ for $\xgo > 0.75$) were not included in these 
$\cos\theta^*$ distributions. The PYTHIA estimation of the contribution of the direct 
process to the resolved-enriched sample, $\xgo < 0.75$, and the resolved process 
to the direct-enriched sample, $\xgo > 0.75$, are also indicated.
    
Direct photon events originating from the dominant $q$-exchange process $\gamma g\to c\bar c$ 
(Fig.~1a)) should have a distribution symmetric in $\cos\theta^*$. The angular 
distribution of direct-enriched events ($\xgo > 0.75$) exhibits a slight asymmetry, 
which can be explained by the feedthrough from resolved photon processes near 
$\cos\theta^* =-1$, as predicted by PYTHIA (Fig.~5b)).               
 
The sample enriched in resolved photons (Fig.~5a,c)) exhibits a mild rise in the proton 
hemisphere towards $\cos\theta^* = 1$, consistent with expectations from quark 
exchange. In contrast, they have a strong rise towards $\cos\theta^* = -1$, i.e. 
in the photon direction, consistent with a dominant contribution from gluon exchange. 
For the latter case, the charm quark emerges in the photon hemisphere (Fig.~1d)). 
Gluon-exchange diagrams with this topology can only come, at LO, from the processes 
$c^{\gamma} g^p\to c g$ and $c^{\gamma} q^p\to c q$, where the superscripts refer 
to an origin in either the photon or proton. The partonic cross sections 
for these $2\to 2$ subprocesses are highly asymmetric in $\cos\theta^*$ and show 
a steep rise towards the photon direction, while the subprocess $gg\to c\bar c$ 
(Fig.~1b)) is symmetric in $\cos\theta^*$. This observation suggests that the 
source of the LO gluon-exchange contribution as seen in Figs.~4a) and c) is charm originating 
from the photon. This is consistent with the MC prediction~\cite{epj:c6:67} that most 
of the resolved photon contribution to charm dijet events at HERA is due to charm 
originating from the photon.                                                  
 
\section{Comparisons with theoretical predictions}
\label{sec-com}
 
\subsection{Comparison with MC predictions}
 
Figures 2--5 compare  the distributions of the data with those of the MC simulations  
PYTHIA, HERWIG and CASCADE. For PYTHIA and HERWIG, the predictions are normalised to the 
data with normalisation factors shown in brackets within the figures.                
For a shape comparison, the prediction
for CASCADE is shown in Fig.~2 normalised to the data.                        
%                                                      Since CASCADE has been 
%previously compared to data with absolute cross section                     
%predictions~\cite{epj:c24:425},                                        
Since there is a hope~\cite{pan:64:120} that higher-order
corrections to $k_t$-factorised calculations might be smaller than
those to LO parton-shower calculations using DGLAP evolution, the absolute predictions
from CASCADE for the differential cross sections are shown
in Figs.~3~--~5.
 
                             The shapes of all
    data distributions are well reproduced by PYTHIA. The HERWIG predictions give 
an adequate description of the shapes in the data, although the rise in the cross
section as a function of $\cos\theta^*$ at low $\xgo$ is stronger in the data,
particularly in the photon direction.
        There is a sizeable contribution from a resolved photon component in both 
PYTHIA ($35\%$) and HERWIG ($22\%$). Fitting the MC distributions to the data, 
allowing the resolved and direct photon contributions to vary independently, results 
in a resolved contribution of $46\%$ for PYTHIA and $30\%$ for HERWIG. The fraction 
of charm dijet events that originates from beauty production is predicted to be 
$\approx 10\%$ by PYTHIA and $\approx 6\%$ by HERWIG. The shape of the beauty 
component is similar to that of the overall distributions.
 
The $\xgo$ distribution of CASCADE, normalised to the data, 
                  gives a larger contribution at high $\xgo$ and a smaller contribution 
at low $\xgo$ (Fig.~2a)).                                           
The absolute cross section predictions for CASCADE, shown in Figs.~3 -- 5,
               are larger than the data by around $30\%$. This difference 
is concentrated in the region $\xgo > 0.75$ and cannot be accounted for by a variation
of $m_c$: changing $m_c$ to 1.3 and 1.7$\gev$ gave a deviation in the prediction of
$\pm 10\%$. However, the CASCADE prediction reproduces the
shape in $\xpo$.
                          The angular                distributions
are well described for $\xgo~<~0.75$, although CASCADE underestimates the
data in the proton direction (Fig.~5c). For $\xgo~>~0.75$ (Fig.~5d), the prediction
overestimates the data in all regions of $\cos\theta^*$, although the shape
is described reasonably well.
%for the sample enriched in resolved photon events, both in the proton and photon 
%directions (Fig.~5a)). For the sample enriched in direct photon events, the 
%$\cos\theta^*$ distribution rises more steeply towards the photon direction in CASCADE
%   than it does in the data (Fig.~5b)).
 
%**************Discuss absolute CASCADE comparison******************************
 
\subsection{Comparison with NLO QCD predictions}
 
The differential cross sections of Figs.~2--5 have been compared to   the NLO FO 
calculation~\cite{np:b412:225,*pl:b348:633,*np:b454:3}. The uncertainties in the NLO calculation, 
shown as the shaded area, come from the simultaneous variation of 
$m_c$  between 1.3 and 1.7$\gev$ and $\mu_R$ between $m_T /2$ and $2m_T$. Changing 
the photon PDF parameterisation from AFGHO to GRVHO~\cite{pr:d45:3986,pr:d46:1973}, 
as well as varying $\mu_F$ of the photon and proton PDFs between 
$m_T /2$ and $2m_T$, produce small effects ($< 5\%$) on the NLO predictions.
 
The differential cross sections predicted by the FO calculation were corrected for 
hadronisation effects. For each bin, the partonic cross section was multiplied by a 
hadronisation correction factor, 
$C_{\rm had}=\sigma_{\rm MC}^{\rm hadrons}/\sigma_{\rm MC}^{\rm partons}$, which is 
the ratio of the MC cross sections after and before the hadronisation process. The 
value of $C_{\rm had}$ was taken as the mean of the ratios obtained using HERWIG 
and PYTHIA. Half the spread between the two MCs was added in quadrature to   the 
uncertainty in the NLO calculation. The deviation of $C_{\rm had}$ from unity is
typically below $20\%$ (see Tables 1-4).
% The deviation of $C_{\rm had}$ from unity is           below 
%$20\%$, except for the second-to-highest $\xgo$ bin  (Fig.~2b))              
% and the highest $|\cos\theta^*|$ (Fig.~4d))                              
% and     lowest $\cos\theta^*$ (Fig.~5d)) bins of the sample enriched 
%in direct photons,                  where it is below $30\%$ and for     
%     the sample enriched in resolved photons       (Fig.~5c)) in the two 
%% highest $\cos\theta^*$ bins, where it is             
%%                                  40 -- 60$\%$. 
 
Figure 2b) shows a comparison for the differential cross section in $\xgo$. To 
minimise the large migration effects at $\xgo > 0.75$ due to hadronisation, a 
wider bin than that of Fig.~2a) was used. Migrations to low $\xgo$ are small.       
%Since the FO calculation~\cite{pl:b348:633,*np:b454:3} has no explicit charm contribution 
%to the photon structure, the Born term has mainly a direct photon part with 
%negligible contributions from the resolved photon processes $gg\to c\bar c$ 
%(Fig.~1b)) and $q\bar{q} \to g \to c\bar c$.                                      
    The cross section can have a low $\xgo$ contribution at NLO due to
three-parton final states in which one of the partons is treated as a
photon remnant. However, the low $\xgo$ tail of the NLO cross section
is below the data~\cite{epj:c6:67}.  For $\xgo > 0.75$, the data are
well described by the NLO prediction.
 
The differential cross section as a function of $\xpo$ is compared in Fig.~3b) with 
the NLO FO calculation.                          The NLO prediction is in 
reasonable agreement with the data. As expected from the $\xgo$ comparison, the NLO 
prediction for the resolved-enriched $\xpo$ distribution (not shown) is too low, but 
the shape is well reproduced.
 
Figures 4c-d) and 5c-d) compare the charm dijet angular distributions to the 
NLO calculation. For high $\xgo$ (Figs.~4d) and 5d)), the NLO prediction gives a 
good description of the data. For low $\xgo$ (Fig.~4c)), the NLO prediction is 
significantly below the data. In Fig.~5c), the NLO predicts a lower cross section 
than the data in both proton and photon directions. The shapes of the 
$|\cos\theta^*|$ and $\cos\theta^*$ distributions are reasonably well described 
by the NLO predictions.
 
%\subsection{Systematic Uncertainties}
% ----------------------------------------------------------------------------
 
\section{Conclusions}
\label{sec-con}
 
The differential cross sections as a function of $\cos\theta^*$ for charm 
dijet photoproduction events~ (~median~$Q^2\approx~3\cdot~10^{-4}$\gev$^2$)~ have~ been 
~measured~ in~ the~ kinematic~ range \\
$130~<~W~<~280\gev$, $Q^2 < 1\gev^2$, $p_T^{D^*} > 3\gev$, $|\eta^{D^*}| < 1.5$, 
$E_T^{\rm jet} > 5\gev$ and $|\eta^{\rm jet}| < 2.4$. The  cuts on the dijet invariant 
mass, $M_{\rm jj} > 18\gev$, and on the average jet pseudorapidity, $|\bar\eta| < 0.7$, 
select an 
      $M_{\rm jj}$ and $|\bar\eta|$ region where the biases from other
kinematic cuts are minimised. The distributions have been measured 
separately for samples of events enriched in resolved ($\xgo < 0.75$) and direct ($\xgo > 0.75$) 
photon processes. The angular dependence for the two samples is significantly 
different, reflecting the different spins of the quark and gluon propagators. The 
cross section rises faster with increasing $|\cos\theta^*|$ for resolved 
photoproduction, where processes involving spin-1 gluon exchange dominate, than for 
direct photoproduction, where processes involving spin-1/2 quark exchange dominate.     
 
The shapes of the measured differential cross sections are well
reproduced by PYTHIA.  Except for the angular distributions at low
$\xgo$, HERWIG gives an adequate description of these shapes.  The
predictions of CASCADE describe the data at low $\xgo$ in both shape
and normalisation. For high $\xgo$, the prediction significantly
overestimates the data, but gives a reasonable description of the
shapes.
%undershoot the data for the resolved photon case and overshoot it for the direct 
%photon case.                                                            
              The shapes of the measured angular distributions   are 
approximately reproduced by the NLO FO predictions. The absolute cross sections
predicted by the 
NLO FO calculation reproduce the data for the sample enriched in direct photons       
but are below the data for the sample enriched in resolved photons.          
 
%******************Summarize absolute CASCADE comparison******************
 
Associating the $D^*$ meson with one of the jets allows the sign of
$\cos\theta^*$ to be defined. In all cases, the $\cos\theta^*$
distributions show a mild rise towards $|\cos\theta^*| =1$, as
expected from quark exchange, except for the resolved-enriched sample
in which the cross section rises steeply in the photon direction
($\cos\theta^* = -1$), as expected from gluon exchange. This
observation indicates that most of the resolved photon contribution in
LO QCD charm production is due to charm originating from the photon,
rather than to the competing resolved photon process $g g \to c \bar
c$. This demonstrates that charm originating from the photon is the
dominant component in the resolved photoproduction of dijet events
with charm.
 
\section{Acknowledgments}
\label{sec-ack}
 
We thank the DESY Directorate for their strong support and encouragement. The
remarkable achievements of the HERA machine group were essential for the successful
completion of this work and are greatly appreciated. We are grateful for the support of
the DESY computing and network services. The design, construction and installation of
the ZEUS detector have been made possible owing to the ingenuity and effort of many
people from DESY and home institutes who are not listed as authors.
We thank H.~Jung for informative discussions and S.~Frixione for providing his
NLO code.
 
% ----------------------------------------------------------------------------
 
\vfill\eject
 
%%% Local Variables: 
%%% mode: latex
%%% TeX-master: t
%%% End: 

%------------------------------------------------------------------------------
%       Bibliography
%------------------------------------------------------------------------------
{
\def\bibname{\Large\bf References}
\def\refname{\Large\bf References}
\pagestyle{plain}
\ifzeusbst
  \bibliographystyle{./BiBTeX/bst/l4z_default}
\fi
\ifzdrftbst
  \bibliographystyle{./BiBTeX/bst/l4z_draft}
\fi
\ifzbstepj
  \bibliographystyle{./BiBTeX/bst/l4z_epj}
\fi
\ifzbstnp
  \bibliographystyle{./BiBTeX/bst/l4z_np}
\fi
\ifzbstpl
  \bibliographystyle{./BiBTeX/bst/l4z_pl}
\fi
{\raggedright
\bibliography{./BiBTeX/user/syn.bib,%
              ./BiBTeX/bib/l4z_articles.bib,%
              ./BiBTeX/bib/l4z_books.bib,%
              ./BiBTeX/bib/l4z_conferences.bib,%
              ./BiBTeX/bib/l4z_h1.bib,%
              ./BiBTeX/bib/l4z_misc.bib,%
              ./BiBTeX/bib/l4z_old.bib,%
              ./BiBTeX/bib/l4z_preprints.bib,%
              ./BiBTeX/bib/l4z_replaced.bib,%
              ./BiBTeX/bib/l4z_temporary.bib,%
              ./BiBTeX/bib/l4z_zeus.bib}}
}
\vfill\eject

%------------------------------------------------------------------------------
%       Tables
%------------------------------------------------------------------------------
\renewcommand{\arraystretch}{1.2}

\begin{table}[hbt]
  \begin{center}
  \begin{tabular}{|c|ccccc||c|} \hline 
   $\xgo$ bin &  $d\sigma/d\xgo$ & $\Delta_{\rm stat}$ & $\Delta_{\rm syst}$ & $\Delta_{\rm ES}$&(nb)& $C_{\rm had}$\\ \hline \hline
   0.250, 0.375 &  0.115 & $\pm  0.029$ & $^{+ 0.037}_{- 0.017}$ & $_{ -0.001}^{+ 0.004}$ & & 0.941 $\pm$ 0.040 \\ %\hline
   0.375, 0.500 &  0.196 & $\pm  0.055$ & $^{+ 0.064}_{- 0.034}$ & $_{ -0.005}^{+ 0.022}$ & & 0.950 $\pm$ 0.004 \\ %\hline
   0.500, 0.625 &  0.407 & $\pm  0.082$ & $^{+ 0.101}_{- 0.086}$ & $_{ -0.029}^{+ 0.032}$ & & 1.006 $\pm$ 0.008 \\ %\hline
   0.625, 0.750 &  1.011 & $\pm  0.102$ & $^{+ 0.073}_{- 0.169}$ & $_{ -0.112}^{+ 0.093}$ & & 1.285 $\pm$ 0.050 \\ %\hline
   0.750, 0.875 &  2.000 & $\pm  0.147$ & $^{+ 0.159}_{- 0.105}$ & $_{ -0.159}^{+ 0.150}$ & &           \\ %\hline
   0.875, 1.000 &  1.727 & $\pm  0.122$ & $^{+ 0.243}_{- 0.080}$ & $_{ -0.089}^{+ 0.052}$ & &         \\ \hline
   0.750, 1.000 &  1.864 & $\pm  0.096$ & $^{+ 0.145}_{- 0.066}$ & $_{ -0.124}^{+ 0.101}$ & & 0.851 $\pm$ 0.041 \\ \hline
  \end{tabular}
\end{center}
\vspace{0.7cm}
\caption[]
{Measured cross sections as a function of $\xgo$. The 
statistical, systematic and jet energy scale ($\Delta_{\rm ES}$) uncertainties are shown 
separately. The multiplicative 
hadronisation correction applied to the NLO prediction is shown in the last column. The uncertainty 
shown for the hadronisation correction is half the spread between the values obtained using the 
HERWIG and PYTHIA models.}
\label{table-cos}
\end{table}

\vspace{1.5cm}

\begin{table}[hbt]
  \begin{center}
  \begin{tabular}{|c|ccccc||c|} \hline 
   $\xpo$ bin &  $d\sigma/d\xpo$ & $\Delta_{\rm stat}$ & $\Delta_{\rm syst}$ & $\Delta_{\rm ES}$&(nb)& $C_{\rm had}$\\ \hline \hline
   0.0055, 0.0110 &  23.28 & $\pm  2.47$ & $^{+ 2.70}_{- 2.58}$ & $_{ -0.95}^{+ 0.62}$ & & 0.799 $\pm$ 0.040 \\ %\hline
   0.0110, 0.0165 &  36.90 & $\pm  2.96$ & $^{+ 1.95}_{- 2.74}$ & $_{ -3.05}^{+ 2.53}$ & & 0.910 $\pm$ 0.031 \\ %\hline
   0.0165, 0.0220 &  30.72 & $\pm  2.57$ & $^{+ 2.91}_{- 2.08}$ & $_{ -2.20}^{+ 1.90}$ & & 0.953 $\pm$ 0.027 \\ %\hline
   0.0220, 0.0275 &  18.55 & $\pm  2.05$ & $^{+ 0.97}_{- 1.18}$ & $_{ -1.49}^{+ 1.61}$ & & 0.985 $\pm$ 0.017 \\ %\hline
   0.0275, 0.0330 &   7.84 & $\pm  1.50$ & $^{+ 0.33}_{- 1.88}$ & $_{ -0.37}^{+ 0.58}$ & & 0.984 $\pm$ 0.019 \\ %\hline
   0.0330, 0.0385 &   3.21 & $\pm  0.77$ & $^{+ 0.56}_{- 0.39}$ & $_{ -0.41}^{+ 0.02}$ & & 1.020 $\pm$ 0.047 \\ %\hline
   0.0385, 0.0440 &   2.37 & $\pm  0.69$ & $^{+ 0.55}_{- 0.31}$ & $_{ -0.02}^{+ 0.41}$ & & 1.022 $\pm$ 0.012 \\ \hline
  \end{tabular}
\end{center}
\vspace{0.7cm}
\caption[]
{Measured cross sections as a function of $\xpo$. For further details, 
see the caption to Table~\ref{table-cos}.}
\end{table}

\newpage
\begin{table}[hbt]
  \begin{center}
  \begin{tabular}{|c|ccccc||c|} \hline 
   $|\cos \theta^* |$ bin &  $d\sigma/d|\cos \theta^*|$ & $\Delta_{\rm stat}$ & $\Delta_{\rm syst}$ & $\Delta_{\rm ES}$&(nb)& $C_{\rm had}$\\ \hline \hline
   \multicolumn{7}{|c|}{$\xgo < 0.75$} \\ \hline
   0.00000, 0.10375 &  0.056 & $\pm  0.034$ & $^{+ 0.022}_{- 0.022}$ & $_{ -0.005}^{+ 0.014}$ & & 1.007 $\pm$ 0.014 \\ %\hline
   0.10375, 0.20750 &  0.040 & $\pm  0.027$ & $^{+ 0.028}_{- 0.010}$ & $_{ -0.003}^{+ 0.011}$ & & 1.099 $\pm$ 0.003 \\ %\hline
   0.20750, 0.31125 &  0.126 & $\pm  0.041$ & $^{+ 0.022}_{- 0.026}$ & $_{ -0.014}^{+ 0.011}$ & & 1.072 $\pm$ 0.026 \\ %\hline
   0.31125, 0.41500 &  0.114 & $\pm  0.032$ & $^{+ 0.032}_{- 0.025}$ & $_{ -0.005}^{+ 0.015}$ & & 1.099 $\pm$ 0.048 \\ %\hline
   0.41500, 0.51875 &  0.280 & $\pm  0.062$ & $^{+ 0.055}_{- 0.051}$ & $_{ -0.021}^{+ 0.027}$ & & 1.107 $\pm$ 0.041 \\ %\hline
   0.51875, 0.62250 &  0.300 & $\pm  0.069$ & $^{+ 0.095}_{- 0.059}$ & $_{ -0.050}^{+ 0.009}$ & & 1.101 $\pm$ 0.029 \\ %\hline
   0.62250, 0.72625 &  0.536 & $\pm  0.088$ & $^{+ 0.031}_{- 0.138}$ & $_{ -0.045}^{+ 0.005}$ & & 1.145 $\pm$ 0.014 \\ %\hline
   0.72625, 0.83000 &  0.732 & $\pm  0.108$ & $^{+ 0.087}_{- 0.155}$ & $_{ -0.036}^{+ 0.053}$ & & 1.115 $\pm$ 0.018 \\ \hline
   \multicolumn{7}{|c|}{$\xgo > 0.75$} \\ \hline
   0.00000, 0.10375 &  0.277 & $\pm  0.055$ & $^{+ 0.049}_{- 0.038}$ & $_{ -0.030}^{+ 0.012}$ & & 0.923 $\pm$ 0.069 \\ %\hline
   0.10375, 0.20750 &  0.401 & $\pm  0.065$ & $^{+ 0.037}_{- 0.064}$ & $_{ -0.030}^{+ 0.044}$ & & 0.919 $\pm$ 0.044 \\ %\hline
   0.20750, 0.31125 &  0.471 & $\pm  0.063$ & $^{+ 0.045}_{- 0.080}$ & $_{ -0.039}^{+ 0.020}$ & & 0.910 $\pm$ 0.052 \\ %\hline
   0.31125, 0.41500 &  0.390 & $\pm  0.070$ & $^{+ 0.055}_{- 0.036}$ & $_{ -0.028}^{+ 0.023}$ & & 0.906 $\pm$ 0.068 \\ %\hline
   0.41500, 0.51875 &  0.584 & $\pm  0.082$ & $^{+ 0.048}_{- 0.047}$ & $_{ -0.034}^{+ 0.024}$ & & 0.876 $\pm$ 0.056 \\ %\hline
   0.51875, 0.62250 &  0.636 & $\pm  0.089$ & $^{+ 0.044}_{- 0.040}$ & $_{ -0.030}^{+ 0.017}$ & & 0.863 $\pm$ 0.061 \\ %\hline
   0.62250, 0.72625 &  0.810 & $\pm  0.098$ & $^{+ 0.094}_{- 0.026}$ & $_{ -0.041}^{+ 0.059}$ & & 0.832 $\pm$ 0.036 \\ %\hline
   0.72625, 0.83000 &  0.922 & $\pm  0.126$ & $^{+ 0.105}_{- 0.090}$ & $_{ -0.046}^{+ 0.026}$ & & 0.756 $\pm$ 0.013 \\ \hline
  \end{tabular}
\end{center}
\vspace{0.7cm}
\caption[]
{Measured cross sections as a function of $|\cos \theta^*|$ for
$\xgo < 0.75$ and $\xgo > 0.75$. For further details, 
see the caption to Table~\ref{table-cos}.}
\end{table}

\newpage

\begin{table}[hbt]
  \begin{center}
  \begin{tabular}{|c|ccccc||c|} \hline 
   $\cos \theta^* $ bin &  $d\sigma/d\cos \theta^*$ & $\Delta_{\rm stat}$ & $\Delta_{\rm syst}$ & $\Delta_{\rm ES}$&(nb)& $C_{\rm had}$\\ \hline \hline
   \multicolumn{7}{|c|}{$\xgo < 0.75$} \\ \hline
   -0.830, -0.664 &  0.471 & $\pm  0.072$ & $^{+ 0.065}_{- 0.077}$ & $_{ -0.023}^{+ 0.034}$ & & 1.063 $\pm$ 0.008 \\ %\hline
   -0.664, -0.498 &  0.198 & $\pm  0.036$ & $^{+ 0.043}_{- 0.025}$ & $_{ -0.018}^{+ 0.006}$ & & 1.065 $\pm$ 0.023 \\ %\hline
   -0.498, -0.332 &  0.111 & $\pm  0.028$ & $^{+ 0.028}_{- 0.014}$ & $_{ -0.007}^{+ 0.012}$ & & 1.084 $\pm$ 0.029 \\ %\hline
   -0.332,  0.000 &  0.032 & $\pm  0.011$ & $^{+ 0.009}_{- 0.010}$ & $_{ -0.003}^{+ 0.006}$ & & 1.056 $\pm$ 0.0004 \\ %\hline
    0.000,  0.332 &  0.043 & $\pm  0.015$ & $^{+ 0.009}_{- 0.007}$ & $_{ -0.004}^{+ 0.006}$ & & 1.105 $\pm$ 0.061 \\ %\hline
    0.332,  0.498 &  0.079 & $\pm  0.024$ & $^{+ 0.015}_{- 0.021}$ & $_{ -0.003}^{+ 0.020}$ & & 1.178 $\pm$ 0.140 \\ %\hline
    0.498,  0.664 &  0.064 & $\pm  0.035$ & $^{+ 0.050}_{- 0.028}$ & $_{ -0.013}^{+ 0.001}$ & & 1.374 $\pm$ 0.215 \\ %\hline
    0.664,  0.830 &  0.148 & $\pm  0.039$ & $^{+ 0.014}_{- 0.038}$ & $_{ -0.014}^{+ 0.001}$ & & 1.608 $\pm$ 0.248 \\ \hline
   \multicolumn{7}{|c|}{$\xgo > 0.75$} \\ \hline
   -0.830, -0.664 &  0.557 & $\pm  0.066$ & $^{+ 0.069}_{- 0.054}$ & $_{ -0.017}^{+ 0.027}$ & & 0.758 $\pm$ 0.014 \\ %\hline
   -0.664, -0.498 &  0.371 & $\pm  0.048$ & $^{+ 0.024}_{- 0.021}$ & $_{ -0.016}^{+ 0.018}$ & & 0.842 $\pm$ 0.041 \\ %\hline
   -0.498, -0.332 &  0.258 & $\pm  0.046$ & $^{+ 0.034}_{- 0.028}$ & $_{ -0.023}^{+ 0.017}$ & & 0.880 $\pm$ 0.053 \\ %\hline
   -0.332,  0.000 &  0.183 & $\pm  0.024$ & $^{+ 0.022}_{- 0.024}$ & $_{ -0.017}^{+ 0.009}$ & & 0.914 $\pm$ 0.048 \\ %\hline
    0.000,  0.332 &  0.198 & $\pm  0.024$ & $^{+ 0.018}_{- 0.018}$ & $_{ -0.017}^{+ 0.013}$ & & 0.922 $\pm$ 0.062 \\ %\hline
    0.332,  0.498 &  0.212 & $\pm  0.035$ & $^{+ 0.029}_{- 0.008}$ & $_{ -0.008}^{+ 0.013}$ & & 0.892 $\pm$ 0.065 \\ %\hline
    0.498,  0.664 &  0.313 & $\pm  0.053$ & $^{+ 0.024}_{- 0.043}$ & $_{ -0.023}^{+ 0.001}$ & & 0.885 $\pm$ 0.076 \\ %\hline
    0.664,  0.830 &  0.308 & $\pm  0.068$ & $^{+ 0.031}_{- 0.023}$ & $_{ -0.019}^{+ 0.014}$ & & 0.831 $\pm$ 0.071 \\ \hline
  \end{tabular}
\end{center}
\vspace{0.7cm}
\caption[]
{Measured cross sections as a function of $\cos \theta^*$ for
$\xgo < 0.75$ and $\xgo > 0.75$.  For further details, 
see the caption to Table~\ref{table-cos}.} 
\end{table}

%%% Local Variables: 
%%% mode: latex
%%% TeX-master: t
%%% End: 

%------------------------------------------------------------------------------
%       Figures
%------------------------------------------------------------------------------
 %-------------------------------------------------------------------------------
%       Results
%-------------------------------------------------------------------------------
\begin{figure}[p]
\begin{center}
\epsfig{file=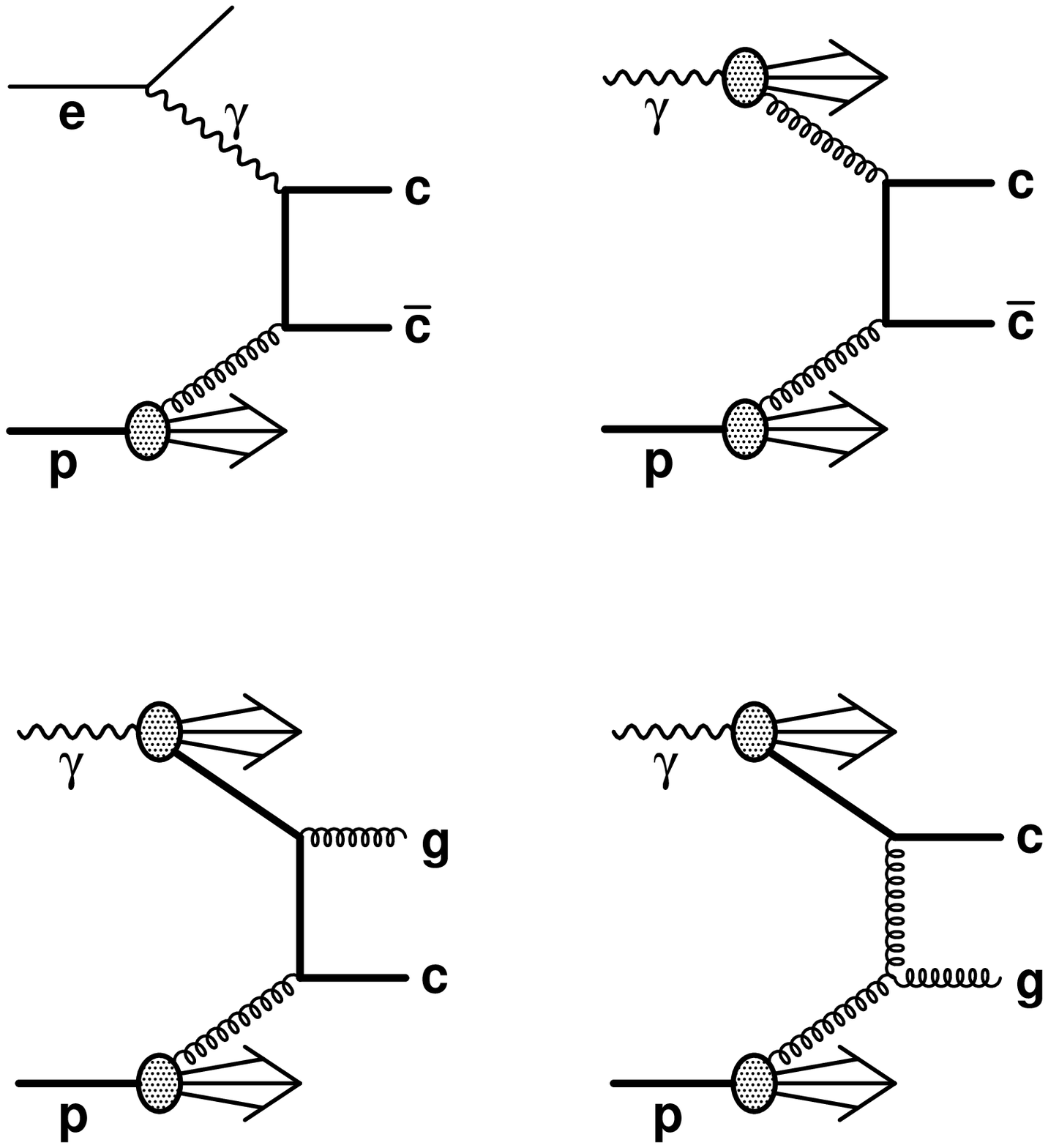,width=16cm}
 \put(-350,400){\Large a)}
 \put(-170,400){\Large b)}
 \put(-350,190){\Large c)}
 \put(-160,190){\Large d)}
\end{center}

\vfill
\vspace{-4cm}
\caption{
LO QCD charm-production diagrams. a) direct photon: $\gamma g\to c \bar c$;
b) resolved photon: $g g\to c \bar c$; c) resolved-photon charm excitation: 
$c g\to g c$ ($c$ in proton hemisphere); d) resolved-photon charm excitation: 
$c g\to c g$ ($c$ in photon hemisphere).
}
\label{fig1}
\vfill
\end{figure}
 
\begin{figure}[p]
 
\begin{center}
 \epsfig{file=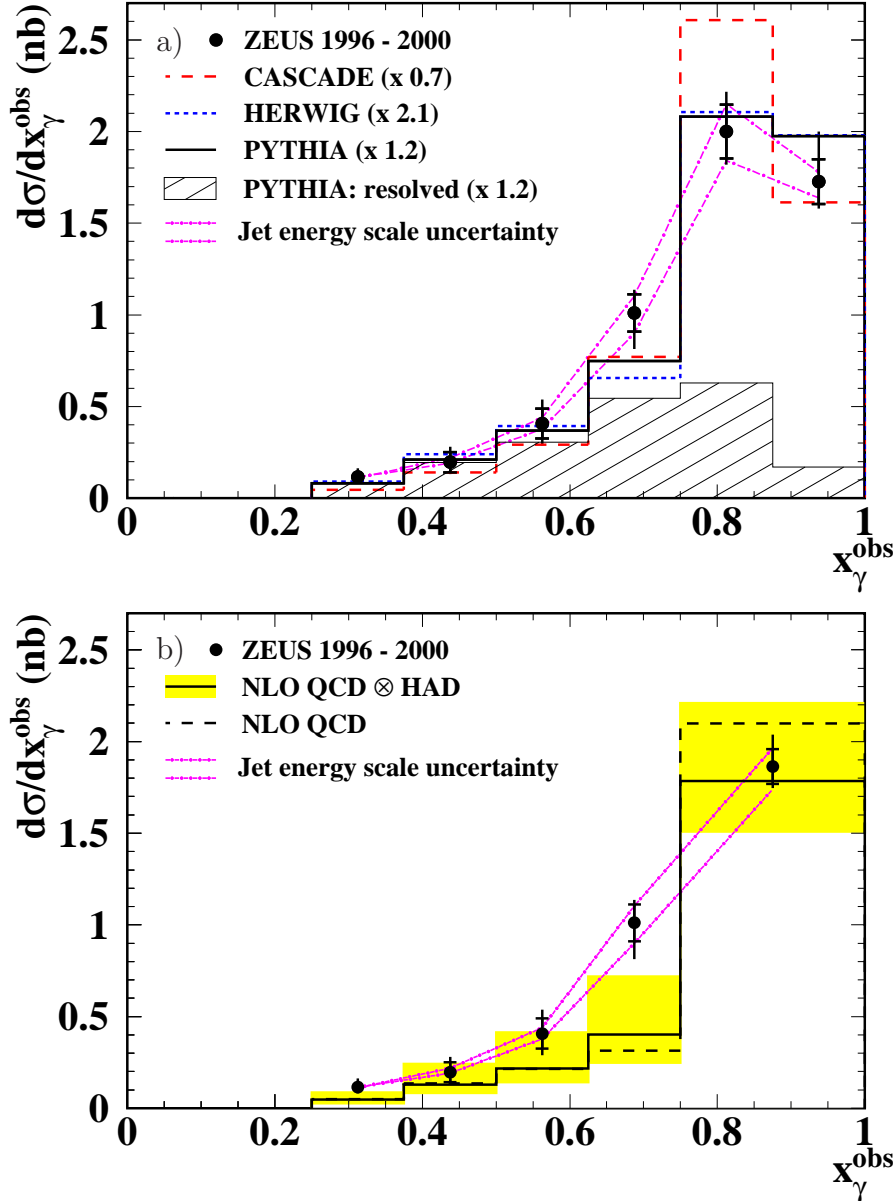,width=12.5cm}
 \put(-280,455){a)}
 \put(-280,225){b)}
\end{center}
\caption{
Differential cross-section $ d\sigma /d\xgo$ for the data (dots)
compared with: a) various MC simulations (histograms);  b) NLO FO predictions after hadronisation 
correction (full lines) and          at parton level (dashed        lines).    
       The inner error bars show the statistical uncertainty, while the outer ones 
show the statistical and systematic uncertainties added in quadrature. The 
jet-energy-scale uncertainty is given by the two dashed-dotted lines. In a), each MC 
distribution is normalised to the data, as indicated in the brackets. Also shown 
in a) is the resolved photon distribution (hatched) of PYTHIA and in b) the 
uncertainty of the NLO prediction after hadronisation correction (shaded). In b) 
the two highest $\xgo$ bins have been combined.
}
\label{fig2}
\end{figure}
 
\begin{figure}[p]
 
\begin{center}
 \epsfig{file=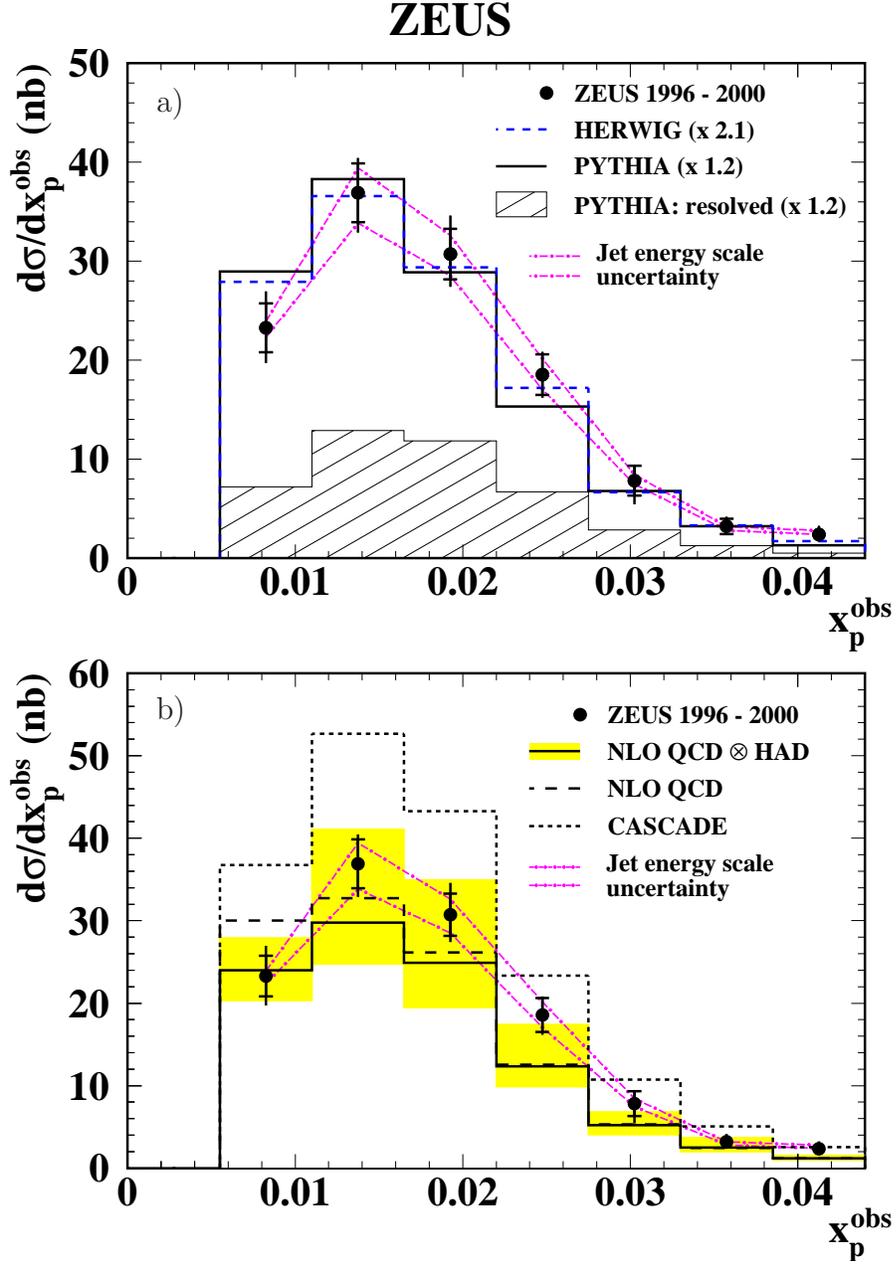,width=12.5cm}
 \put(-280,455){a)}
 \put(-280,225){b)}
\end{center}
\caption{
Differential cross-section $d\sigma /d\xpo$ for the data (dots)
compared with: a) PYTHIA and HERWIG MC simulations (histograms); b) CASCADE (short-dashed lines)
                  and     NLO FO predictions                     
                                                        after hadronisation 
correction (full lines) and          at parton level (long-dashed lines).
The inner error bars show the statistical uncertainty, while the outer ones show 
the statistical and systematic uncertainties added in quadrature. The 
jet-energy-scale uncertainty is given by the two dashed-dotted lines. In a), each MC 
distribution is normalised to the data, as indicated in the brackets. Also shown 
in a) is  the resolved  photon distribution (hatched) of PYTHIA and in b) the 
uncertainty of the NLO prediction after hadronisation correction (shaded).
}
\label{fig3}
\end{figure}
 
\begin{figure}[p]
 
\begin{center}
 \epsfig{file=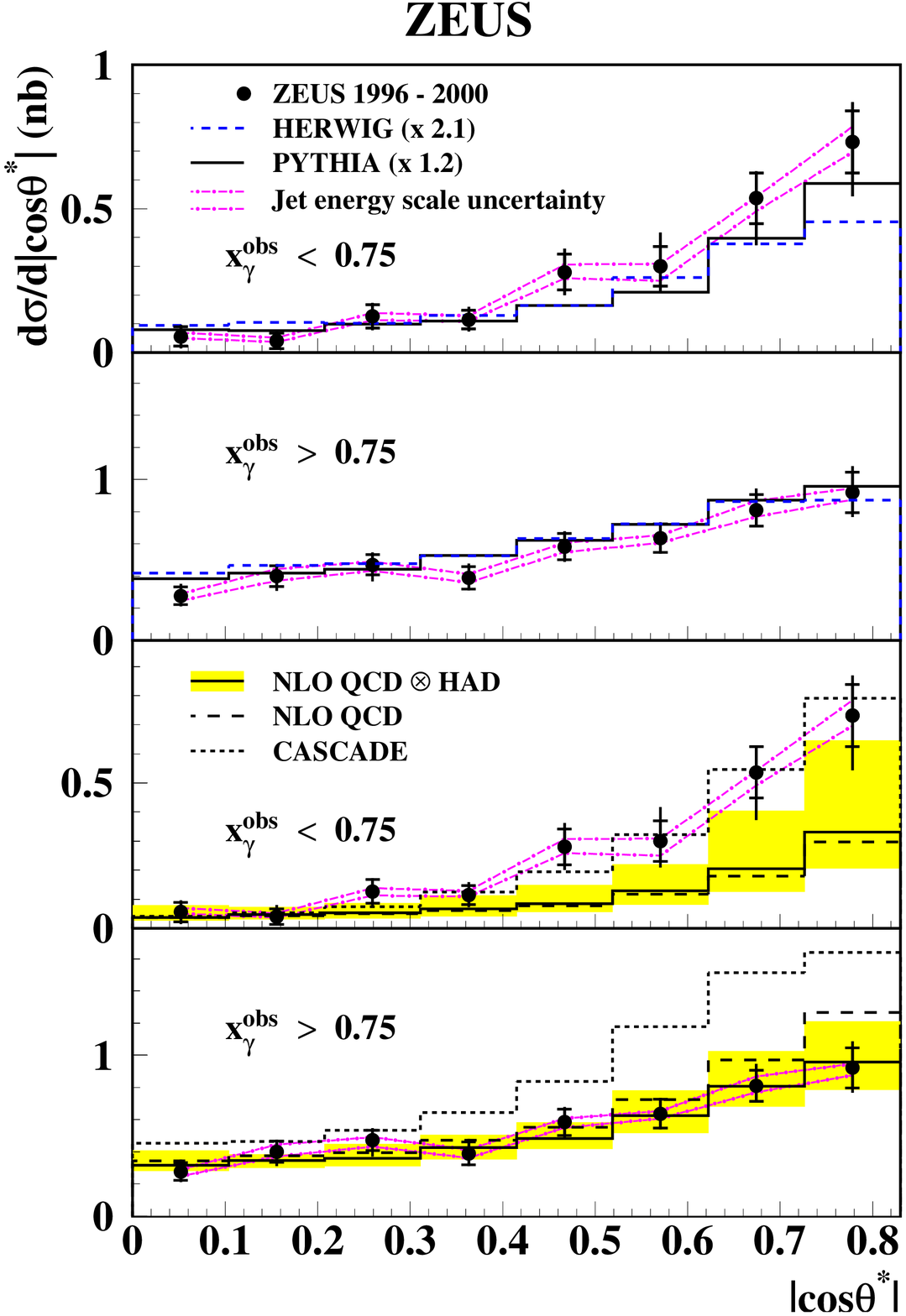,width=12.5cm}
 \put(-280,460){a)}
 \put(-280,353){b)}
 \put(-280,250){c)}
 \put(-280,145){d)}
\end{center}
\caption{          Differential cross-sections $ d\sigma /d|\cos\theta^*|$  
                                    (dots)                               
compared with:  a-b) PYTHIA and HERWIG MC 
simulations (histograms);  c-d) CASCADE (short-dashed lines) and
      NLO FO predictions after hadronisation correction (full lines) and at  
parton level (long-dashed lines).           Results are given separately
in  a,c) for samples enriched in resolved photon events                 and in b,d) for
samples enriched in direct photon events.                  
The inner error bars show the statistical uncertainty, while the outer ones
show the statistical and systematic uncertainties added in quadrature.
The jet-energy-scale uncertainty is given by the two dashed-dotted lines. In a-b), each MC
distribution is normalised to the data, as indicated in the brackets.
Also shown                             
in  c-d) are the uncertainties of the NLO prediction            
          after hadronisation correction             (shaded).  
        }
\label{fig4}
 \end{figure}
 
\begin{figure}[p]
 
\begin{center}
 \epsfig{file=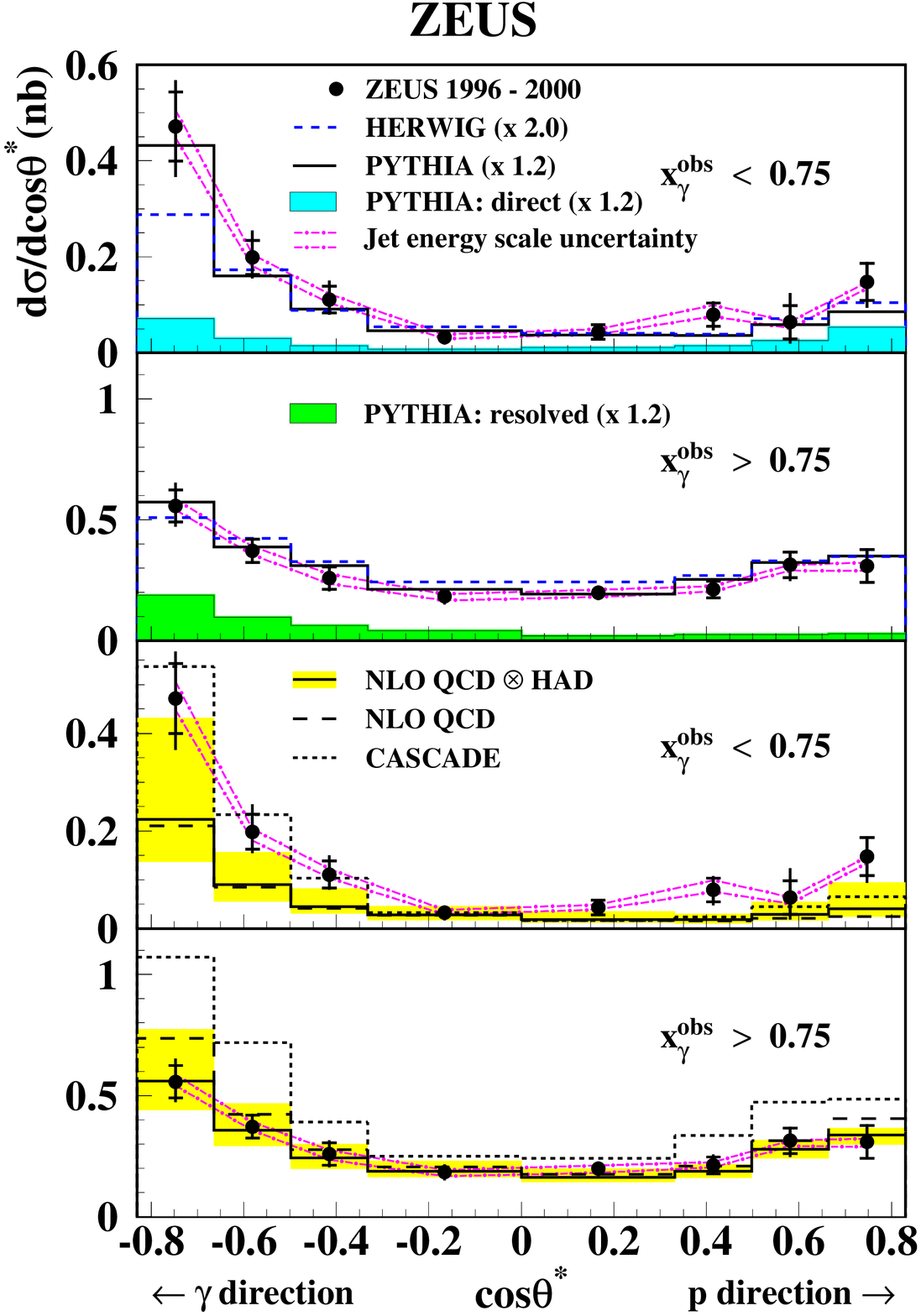,width=12.0cm}
 \put(-50,435){a)}
 \put(-50,332){b)}
 \put(-50,232){c)}
 \put(-50,130){d)}
\end{center}
\vspace{-0.6cm}
\caption{          Differential cross-sections $ d\sigma /d\cos\theta^*$  
                                    (dots)                               
compared with:  a-b) PYTHIA and HERWIG MC 
simulations (histograms);  c-d) CASCADE (short-dashed lines) and           
       NLO FO predictions after hadronisation correction (full lines)
       and at parton level (long-dashed lines). Results are given separately
in  a,c) for samples enriched in resolved photon events                 and in b,d) for 
samples enriched in direct photon events.
The inner error bars show the statistical uncertainty, while the outer ones
show the statistical and systematic uncertainties added in quadrature.
The jet-energy-scale uncertainty is given by the two dashed-dotted lines. In a-b), each MC
distribution is normalised to the data, as indicated in the brackets.
Also shown as shaded areas in a) and b) are the contribution of the
direct photon process in PYTHIA to the resolved-enriched sample and the
contribution of
the resolved photon process to the direct-enriched sample, respectively. The
uncertainties 
of the NLO prediction after the hadronisation correction are shown as the
shaded areas in c) and d).
 }
\label{fig5}
 \end{figure}

%%% Local Variables: 
%%% mode: latex
%%% TeX-master: t
%%% End: 

%
%       ... that's it
%
\end{document}